\documentclass[aps,prl,twocolumn,superscriptaddress,floatfix,showpacs]{revtex4}

\usepackage[T1]{fontenc}
\usepackage{mathptmx}
\usepackage[mathscr]{eucal}
\usepackage{amsfonts}
\usepackage[colorlinks=true,citecolor=blue,linkcolor=blue,urlcolor=blue]{hyperref}
\usepackage{amsmath,amssymb,amsthm,bm}
\usepackage{color}
\usepackage{graphics}
\usepackage{epsfig}
\usepackage{psfrag}
\usepackage{graphicx}

\include{Qcircuit}

\def\la{\langle}
\def\ra{\rangle}

\def\tr{\text{Tr}}
\newcommand{\ket}[1]{|#1\rangle}
\newcommand{\bra}[1]{\langle#1|}

\def\be{\begin{equation}}
\def\ee{\end{equation}}
\def\ba{\begin{eqnarray}}
\def\ea{\end{eqnarray}}

\begin{document}

\title{General optimality of the Heisenberg limit for quantum metrology}
\author{Marcin Zwierz}\affiliation{Department of Physics and Astronomy, University of Sheffield, Hounsfield Road, Sheffield, S3 7RH, UK}
\author{Carlos A.~P\'erez-Delgado}\affiliation{Department of Physics and Astronomy, University of Sheffield, Hounsfield Road, Sheffield, S3 7RH, UK}\affiliation{Department of Physics and Astronomy, University of Sussex, Falmer, Brighton, East Sussex, BN1 9QH, UK}
\author{Pieter Kok}\email{p.kok@sheffield.ac.uk}
\affiliation{Department of Physics and Astronomy, University of Sheffield, Hounsfield Road, Sheffield, S3 7RH, UK}


\begin{abstract}\noindent
Quantum metrology promises improved sensitivity in parameter estimation over classical procedures. However, there is an extensive debate over the question how the sensitivity scales with the resources (such as the average photon number) and number of queries that are used in estimation procedures. Here, we reconcile the physical definition of the relevant resources used in parameter estimation with the information-theoretical scaling in terms of the query complexity of a quantum network. This leads to a completely general optimality proof of the Heisenberg limit for quantum metrology. We give an example how our proof resolves paradoxes that suggest sensitivities beyond the Heisenberg limit, and we show that the Heisenberg limit is an information-theoretic interpretation of the Margolus-Levitin bound, rather than Heisenberg's uncertainty relation. 
\end{abstract}

\maketitle

\noindent
Parameter estimation is a fundamental pillar of science and technology, and improved measurement techniques for parameter estimation have often led to scientific breakthroughs and technological advancement. Caves \cite{Caves81} showed that quantum mechanical systems can in principle produce greater sensitivity over classical methods, and many quantum parameter estimation protocols have been proposed since \cite{kok10}. The field of quantum metrology started with the work of Helstrom \cite{helstrom67,helstrom76}, who derived the minimum value for the mean square error in a parameter in terms of the density matrix of the quantum system and a measurement procedure. This was a generalisation of a known result in classical parameter estimation, called the Cram\'er-Rao bound. Braunstein and Caves \cite{braunstein94} showed how this bound can be formulated for the most general state preparation and measurement procedures. While it is generally a hard problem to show that the Cram\'er-Rao bound can be attained in a given setup, at least it gives an upper limit to the precision of quantum parameter estimation. 

The quantum Cram\'er-Rao bound is typically formulated in terms of the Fisher information, an abstract quantity that measures the maximum information about a parameter $\varphi$ that can be extracted from a given measurement procedure. One of the central questions in quantum metrology is how the Fisher information scales with the physical resources used in the measurement procedure. We usually consider two scaling regimes: First, in the {\em standard quantum limit} ({\sc sql}) \cite{gardiner04} or {\em shot-noise limit} the Fisher information is constant, and the error scales with the inverse square root of the number of times $T$ we make a measurement. Second, in the {\em Heisenberg limit} \cite{holland93} the error is bounded by the inverse of the physical resources. Typically, these are expressed in terms of the size $N$ of the probe system, e.g., (average) photon number. However, it has been clearly demonstrated that this form of the limit is not universally valid. For example,  Beltr\'an and Luis \cite{luis05} showed that the use of classical optical nonlinearities can lead to an error with average photon number scaling $N^{-3/2}$. Boixo {\em et al}.\ \cite{boixo} devised a parameter estimation procedure that sees the error scale with $N^{-k}$ with $k\in\mathbb{N}$, and Roy and Braunstein \cite{Roy08} construct a procedure that achieves an error that scales with $2^{-N}$. The central question is then: What is the real fundamental Heisenberg limit for quantum metrology? We could redefine this limit accordingly to scale as $2^{-N}$, but in practice this bound will never be tight. 

In this Letter, we give a natural definition of the relevant physical resources for quantum metrology based on the general description of a parameter estimation procedure, and we prove the fundamental bound on the mean squared error based on this resource count. We will show that the resource count is proportional to the size of the probe system only if the interaction between the object and the probe is non-entangling over the systems constituting the probe. First, we study the query complexity of quantum metrology networks, which will lead to a resource count given by the expectation value of the generator of translations in the parameter $\varphi$. Second, we prove that the mean error in $\varphi$ is bounded by the inverse of this resource count. We argue that this is the fundamental Heisenberg limit for quantum metrology. Furthermore, we show that it is a form of the Margolus-Levitin bound, as opposed to Heisenberg's uncertainty relation. Finally, we illustrate how this general principle can resolve paradoxical situations in which the Heisenberg limit seems to be surpassed.

\begin{figure}[t]
\includegraphics[width=8cm]{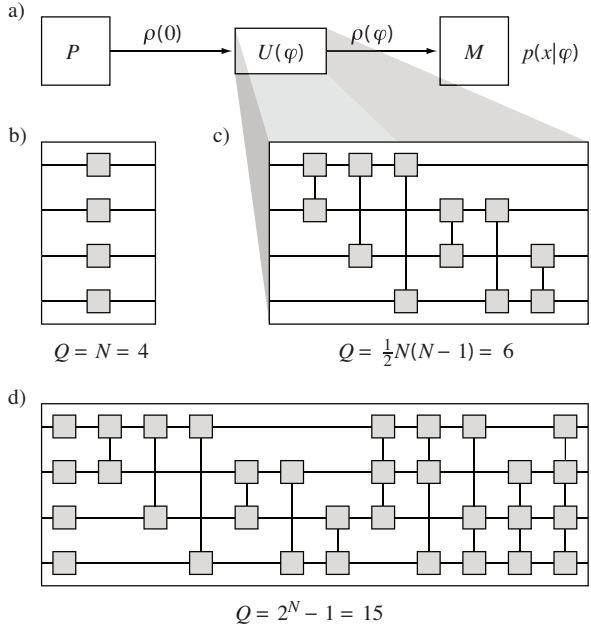}
\caption{\label{fig:setup} a) General parameter estimation procedure involving state preparation $P$, evolution $U(\varphi)$ and generalized measurement $M$ with outcomes $x$, which produces a probability distribution $p(x|\varphi)$. In terms of quantum networks, the evolution can be written as a number of queries of the parameter $\varphi$. b) Example for $N=4$ of the usual situation described by $\mathcal{H}_{\rm GLM}$, where each system performs a single query, and the number of queries equals the number of systems (the grey box represents $O_j(\varphi)$); c) for $\mathcal{H}_{\rm BFCG}$ the number of queries $Q$ does not always equal the number of systems:  any two systems can jointly perform a single query, and the number of queries then scales quadratically with the number of systems; d) for $\mathcal{H}_{\rm RB}$ all possible subsets of systems perform a single query. The number of queries scales exponentially with the number of systems.}
\end{figure}

The most general parameter estimation procedure is shown in Fig.~\ref{fig:setup}a). Consider a probe system prepared in an initial quantum state $\rho(0)$ that is evolved to a state $\rho(\varphi)$ by $U(\varphi)=\exp(-i\varphi\mathcal{H})$. This is a unitary evolution when we include the relevant environment into our description, and it includes feed-forward procedures. The Hermitian operator $\mathcal{H}$ is the generator of translations in $\varphi$, the parameter we wish to estimate. The system is subjected to a generalized measurement $M$, described by a Positive Operator Valued Measure ({\sc povm}) that consists of elements $\hat{E}_x$, where $x$ denotes the measurement outcome. These can be discrete or continuous (or a mixture of both). The probability distribution that describes the measurement data is given by the Born rule $p(x|\varphi) = \tr[\hat{E}_x\,\rho(\varphi)]$, and the maximum amount of information about $\varphi$ that can be extracted from this measurement is given by the Fisher information
\be
 F(\varphi) = \int dx\, \frac{1}{p(x|\varphi)} \left( \frac{\partial p(x|\varphi)}{\partial \varphi} \right)^2\, .
\ee
This leads to the quantum Cram\'er-Rao bound \cite{helstrom67,braunstein94}
\be\label{eq:cr}
 \delta\varphi \geq \frac{1}{\sqrt{T F(\varphi)}}\, ,
\ee
where $(\delta\varphi)^2$ is the mean square error in the parameter $\varphi$, and $T$ is the number of times the procedure is repeated. The {\sc sql} is obtained when the Fisher information is a constant with respect to $T$, and the Heisenberg limit is obtained in a single-shot experiment ($T=1$) when the Fisher information scales quadratically with the resource count. The {\sc sql} and the Heisenberg limit therefore relate to two fundamentally different quantities, $T$ and $F$, respectively. We need to reconcile the meaning of these two limits if we want to compare them in a meaningful way.

To solve this problem, we can define an unambiguous resource count for parameter estimation by recognising that a quantum parameter estimation protocol can be written as a quantum network acting on a set of quantum systems, with repeated ``black-box'' couplings of the network to the system we wish to probe for the parameter $\varphi$ \cite{Giovannetti06}. A black-box is a function that can be univariate or multi-variate. When the function is multi-variate, a {\em query} to the black-box must take the form of multiple input parameters. Likewise, when the operator that describes the fundamental ``atomic'' interaction between the queried system and the probe is a two-body interaction on the probe, then a query can consist only of precisely two input bodies. The scaling of the error in $\varphi$ is then determined by the {\em query complexity} of the network. The number of queries $Q$ is not always identical to the number of physical systems $N$ in the network.

In Fig.~\ref{fig:setup}b-d) we consider three examples. The quantum network with univariate black-boxes in b) was analysed by Giovannetti, Lloyd, and Maccone \cite{Giovannetti06}. Suppose that each grey box in Fig.~\ref{fig:setup} is a unitary gate $O_j(\varphi) = \exp(-i\varphi H_j)$, where $j = 1,\ldots,N$  denotes the system, and $H_j$ is a positive Hermitian operator. It is convenient to define the generator of the joint queries as $\mathcal{H}_{\rm GLM} = \sum_j H_j$, because all $H_j$ commute with each other. The number of queries $Q$ is then equal to the number of terms in $\mathcal{H}_{\rm GLM}$, or $Q=N$. In Fig.~\ref{fig:setup}c) the black-box is bi-variate. This is a type of Hamiltonian considered by Boixo, Flammia, Caves, and Geremia \cite{boixo}, and takes the form $\mathcal{H}_{BFCG} = \sum_{k=1}^N \sum_{j=1}^{k}  H_j\otimes H_k$. A physical query to a black-box characterized by $O_{jk} = \exp(-i\varphi H_j \otimes H_k)$ must consist of two systems, labeled $j$ and $k$. Since each pair interaction is a single query, the total number of queries is $\binom{N}{2} = \frac12 N(N-1)$. Finally, in Fig.~\ref{fig:setup}d) we depict the network corresponding to the protocol by Roy and Braunstein \cite{Roy08}. It is easy to see that the number of terms in the corresponding generator $\mathcal{H}_{RB}$ is given by $2^N-1$, and the number of queries is therefore $Q=2^{N}-1$. 

A similar argument can be made to find the correct number of queries for all types of networks. The key principle is that a physical query to a quantum system consists of probe-systems that \emph{together} undergo an  operation, which can potentially entangle them. The entangling power of the black-box operation over multiple input systems  accounts for the super-linear scaling of $Q$ with $N$. Only when $\mathcal{H}$ does not have any entangling power across the input, we are guaranteed to have $Q=O(N)$. This is in agreement with Refs.~\cite{boixo} and \cite{Roy08} where $\sqrt{F(\varphi)}$ scales super-linearly in $N$, but is always linear in $Q$, as defined here. Since we have a systematic method for increasing $N$ (and $Q$) given the atomic interaction $H_j$, this uniquely defines an asymptotic query complexity of the network. Since both $T$ and $Q$ count the number of queries, this allows us to meaningfully compare the {\sc sql} with the Heisenberg limit. 

Given that in Eq.~(\ref{eq:cr}) $\sqrt{F(\varphi)}\lesssim Q$, we have to find a general procedure that bounds $Q$, based on the physical description of the estimation protocol in Fig.~\ref{fig:setup}a). Previously, we showed that $Q$ is the number of black-box terms in $\mathcal{H}$, and a straightforward choice for the resource count is therefore $|\la\mathcal{H}\ra| \leq O(Q)$. An important subtlety occurs when $\mathcal{H}$ corresponds to a proper Hamiltonian. The origin of the energy scale has no physical meaning, and the actual value of $|\la \mathcal{H}\ra|$ can be changed arbitrarily. Hence, we must fix the scale such that the ground state has zero energy (equivalently, we may choose $|\la\mathcal{H}-h_{\rm min}I\ra|$, where $h_{\rm min}$ is the smallest eigenvalue, and $I$ the identity operator). In most cases, this is an intuitive choice. For example, it is natural to associate zero energy to the vacuum state, and add the corresponding amount of energy for each added photon. Technically, this corresponds to the normal ordering of the Hamiltonian of the radiation field in order to remove the infinite vacuum energy. Slightly less intuitive is that the average energy of $N$ spins in a GHZ  state $(\ket{\!\uparrow}^{\otimes N} + \ket{\!\downarrow}^{\otimes N})/\sqrt{2}$ is no longer taken to be zero, but rather $N/2$ times the energy splitting between $\ket{\!\uparrow}$ and $\ket{\!\downarrow}$.

While the expectation value of $\mathcal{H}$ is easy to calculate, it is not the only way to obtain a bound of $O(Q)$ from $\mathcal{H}$. Other seemingly natural choices are the variance and the semi-norm. For example, if we write $\mathcal{H} \equiv \sum_j^Q A_j$, the variance is 
$ (\Delta\mathcal{H})^2 = \la(\sum_j^{Q} A_j)^2\ra - \la \sum_j^Q A_j\ra^2  =  \sum_j^{Q^2} \la L_j \ra - \sum_{j,k}^Q \la A_j\ra\la A_k\ra \leq c Q^2$
for some positive number $c$ and positive operator $L_j$. This gives $\Delta\mathcal{H}\leq O(Q)$, where e.g., in Ref.~ \cite{boixo} $Q=O(N^2)$. Similarly, $|\la\mathcal{H}\ra| = \sum_j^Q |\la A_j\ra| \leq O(Q)$ since all expectation values are positive and finite. In other words, in terms of the scaling behaviour with $Q$, we can use either the variance or the expectation value. However, there are important classes of quantum systems for which the variance of the energy diverges, such as systems with a Breit-Wigner (or Lorentzian) spectrum \cite{breit,uffink93}. Similarly, the semi-norm does not exist for a large class of states, such as optical Gaussian states. In these cases the resource count, and by implication the scaling of the error, would be ill-defined. By contrast, $|\la\mathcal{H}\ra|$ always exists and is always positive. Also, from a physical perspective the higher-order moments do not describe ``amounts'' in the same way as the first moment does, and refer instead to the shape of the distribution. This is a further argument that $|\la\mathcal{H}\ra|$ is the natural choice for the resource count. Sometimes, it is unclear how the query complexity is defined, for example when the estimation procedure does not involve repeated applications of the gates $O_j(\varphi)$, or when an indeterminate number of identical particles, such as photons, are involved. Nevertheless, the generator $\mathcal{H}$ is always well-defined in any estimation procedure, and we can always use its expectation value to define the relevant resource count.

The resource count in terms of $|\la\mathcal{H}\ra|$ is completely general for all possible quantum networks. For interactions $U(\varphi)$ where we include feed-forward and arbitrary unitary gates between queries, we can use an argument by Giovannetti {\em et al}.\ \cite{Giovannetti06} to show that $|\la\mathcal{H}\ra| = |\la (\partial U(\varphi)/\partial\varphi)U^{\dagger}(\varphi)\ra|$ is unaffected by the intermediate unitary gates, and the scaling is therefore still determined by $Q$.

After establishing the appropriate resource count, we are finally in a position to prove the optimality of the Heisenberg limit for quantum parameter estimation in its most general form.  The Fisher information can be related to a statistical distance $s$ on the probability simplex spanned by $p(x|\varphi)$. Consider two probability distributions $p(x|\varphi)$ and $p(x|\varphi) + dp(x)$. The infinitesimal statistical distance between these distributions is given by $ds^2 = \int dx\, [dp(x|\varphi)]^2/p(x|\varphi)$ \cite{bhatta43,wootters81}. Dividing both sides by $(d\varphi)^2$ we obtain
\be\label{eq:sf}
 \left( \frac{ds}{d\varphi} \right)^2 = \int dx\, \frac{1}{p(x|\varphi)} \left( \frac{\partial p(x|\varphi)}{\partial \varphi} \right)^2 = F(\varphi)\, ,
\ee
which relates the Fisher information to the rate of change of the statistical distance (i.e., the speed of dynamical evolution).

When we count the resources that are used in a parameter estimation procedure, we must make sure that we do not leave anything out, and this can be guaranteed by including in our description the environment that the estimation procedure couples to. This reduces the quantum states to pure states, which means that we can use Wootters' distance \cite{wootters81} between quantum states as the statistical distance:
\be\label{eq:woot}
 s(\psi,\phi) = \arccos(|\la\psi|\phi\ra|)\, ,
\ee
where $\ket{\psi}$ and $\ket{\phi}$ are two pure states in the larger Hilbert space, and $s(\psi,\phi)$ is the angle between them. The distance between the probe state $\rho(0)$ and the evolved state $\rho(\varphi)$ can then be represented by the pure states $\ket{\psi(0)}$ and $\ket{\psi(\varphi)}$, respectively, and the unitary evolution is given by 
\be \label{eq:evo}
\ket{ \psi(\varphi)} = \exp\left(-i\varphi\mathcal{H}\right)\ket{ \psi(0)}\, .
\ee
Here, {\em we place no restriction} on $\mathcal{H}$, other than fixing the energy scale if necessary.
We can place an upper bound on the derivative of Wootters' distance by evaluating the differential of $s$ in Eq.~(\ref{eq:woot}) and using the Schr\"odinger equation implicit in Eq.~(\ref{eq:evo}) \cite{jones10}:
\be\label{eq:ml}
 \frac{ds}{d\varphi} \leq |\la\mathcal{H}\ra|\, .
\ee 
Combining this with Eq.~(\ref{eq:sf}) and Eq.~(\ref{eq:cr}) leads to the Cram\'er-Rao bound 
\be\label{eq:kok}
 (\delta\varphi)^2 \geq \frac{1}{T} \left( \frac{ds}{d\varphi} \right)^{-2} \geq \frac{1}{T\, |\la\mathcal{H}\ra|^2}\, .
\ee
When all resources are used in a single-shot ($T=1$) experiment, the error in $\varphi$ is bounded by
\be\label{eq:hl}
 \delta\varphi \geq \frac{1}{|\la\mathcal{H}\ra|}\, .
\ee
Since $|\la\mathcal{H}\ra|$ is the resource count in the parameter estimation procedure, this is the Heisenberg limit. It is always positive and finite, and in the limit where $|\la\mathcal{H}\ra| \to 0$ there are no resources available to estimate $\varphi$, and $\delta\varphi$ cannot be bounded. In general, the bound is not tight. Indeed, only carefully chosen entangled systems can achieve this bound \cite{Giovannetti06}. This completes the proof of the optimality of the Heisenberg limit in the most general case.

In addition to Eqs.~(\ref{eq:sf}) and (\ref{eq:ml}), the Fisher information is also bounded by the variance of $\mathcal{H}$ according to $F(\varphi) \leq 4 (\Delta\mathcal{H})^2$ \cite{braunstein96b}. This leads to a (single-shot) quantum Cram\'er-Rao bound
\be\label{eq:up}
 \delta\varphi \geq \frac{1}{2\Delta\mathcal{H}}\, .
\ee
However, since $\Delta\mathcal{H}$ is not a resource count, such as the average photon number, but rather a variance (or uncertainty) this is not the Heisenberg limit. In fact, it is Heisenberg's uncertainty relation for the parameter $\varphi$ and its conjugate operator $\mathcal{H}$. Any parameter estimation procedure must respect both bounds, and the Heisenberg limit in Eq.~(\ref{eq:hl}) may not be attained for a particular input state because the bound in Eq.~(\ref{eq:up}) prevents it from doing so. 

The term ``Heisenberg limit'' was introduced by Holland and Burnett \cite{holland93}, who referred to the number-phase uncertainty relation in Heitler \cite{heitler54}. However, as our optimality proof and the subsequent discussion indicate, the Heisenberg limit is {\em not} an uncertainty relation, since it relates the uncertainty of the parameter to the {\em first} moment of the conjugate observable $\mathcal{H}$, rather than the second. It turns out instead that the Heisenberg limit is intimately connected to the Margolus-Levitin bound on the time it takes for a quantum system to evolve to an orthogonal state \cite{margolus98,jones10,giovannetti03}. To see this, we can formally solve Eq.~(\ref{eq:ml}) by separation of variables, yielding 
\be
 \int_0^{\varphi} d\varphi' \geq \frac{1}{|\la\mathcal{H}\ra|} \int_0^{\pi/2} ds \quad\Rightarrow\quad \varphi \geq \frac{\pi}{2} \frac{1}{|\la\mathcal{H}\ra|}\, .
\ee
We can therefore identify the Heisenberg limit with the Margolus-Levitin bound on the speed of dynamical evolution of quantum systems when $\mathcal{H}$ is the Hamiltonian. The (generalized) uncertainty relation, on the other hand, can be identified with the Mandelstam-Tamm bound \cite{jones10}. Both limits are completely general (assuming the existence of $\Delta\mathcal{H}$) and complement each other. 

Finally, we demonstrate that our proof applies to continuous variable systems as well as discrete systems, by considering the procedure of Beltr\'an and Luis \cite{luis05}. The construction is as follows: The evolution $O(\varphi)$ is generated by an optical nonlinearity proportional to the square of the photon number operator $\hat{n}^2$ acting on a single-mode coherent state $\ket{\psi(0)}=\ket{\alpha}$. The evolved state before detection is given by $\ket{\psi({\varphi})} = \exp(-i\varphi\hat{n}^2)\ket{\alpha}$, and the mean square error in $\varphi$ is calculated as $\delta\varphi \simeq \frac14 \la \hat{n}\ra^{-3/2} = \frac14 |\alpha|^{-3}$, to leading order in the average photon number $\la \hat{n}\ra$. Since here the average energy is directly proportional to the average photon number, this procedure seems to surpass the Heisenberg limit. To resolve this paradox, we note that the generator of translations in $\varphi$ is {\em not} the photon number operator $\hat{n}$, but rather the higher-order nonlinearity $\mathcal{H} = \hat{n}^2$. The appropriate resource count is therefore $|\la\mathcal{H}\ra| = \la \hat{n}^2\ra$, instead of the average photon number $\la\hat{n}\ra$. It is easily verified that to leading order $\delta\varphi$ is theoretically bounded by $1/\la\hat{n}^2\ra = 1/|\alpha|^4$. Hence the parameter estimation procedure not only does {\em not} beat the Heisenberg limit, it does not attain it.

Formally, we can attain the Heisenberg limit in this setup with the following modification of the input state and the measurement. Consider the single-mode input state $|\psi_{0}\rangle = \left(|0\rangle + |N\rangle\right)/\sqrt{2}$, where $\ket{0}$ denotes no photons, and $\ket{N}$ denotes $N$ photons. The state of the probe before detection is then given by $|\psi(\varphi)\rangle = \exp(-i\varphi \hat{n}^2) |\psi(0)\rangle = (\ket{0} + e^{-i\varphi N^2}\ket{N})/\sqrt{2}$. We define the measurement observable $X = \ket{0}\bra{N}+ \ket{N}\bra{0}$. Hence, for the final state $\ket{\psi(\varphi)}$ we calculate $\langle X \rangle = \langle \psi_{\varphi}| X |\psi_{\varphi}\rangle = \mbox{cos}(N^2 \varphi)$ and $\Delta X = \mbox{sin}(N^2 \varphi)$. Using the standard expression for the mean squared error, we find that 
\begin{equation}
 \delta \varphi = \frac{\Delta X}{\left| d\langle X \rangle/d\varphi \right|} = \frac{1}{N^2}\, .
\end{equation}
Since $|\la\mathcal{H}\ra| = \la \hat{n}^2\ra = \frac12 N^2$, this attains the Heisenberg limit. This is a formal demonstration that the Heisenberg limit can be attained according to quantum mechanics, even though we currently do not know how to implement it. 

In conclusion, we demonstrated that the Heisenberg limit is optimal for all parameter estimation procedures in quantum metrology, but it requires careful consideration as to which resource is appropriate for expressing the scaling behaviour of the mean square error. The correct resource to take into account is (the expectation value of) the generator of the translations in the parameter. In the case of most optical phase estimation protocols this reduces to the average photon number. Contrary to the origin of the term ``Heisenberg limit'', it is not a generalised uncertainty relation, but rather an expression of the Margolus-Levitin bound on the speed of dynamical evolution for quantum states.

\paragraph{Acknowledgements.}
We thank Jonathan Dowling for establishing the etymology of the term ``Heisenberg limit'', and Sam Braunstein for valuable comments on the manuscript. This research was funded by the White Rose Foundation.


\begin{thebibliography}{18}
\expandafter\ifx\csname natexlab\endcsname\relax\def\natexlab#1{#1}\fi
\expandafter\ifx\csname bibnamefont\endcsname\relax
  \def\bibnamefont#1{#1}\fi
\expandafter\ifx\csname bibfnamefont\endcsname\relax
  \def\bibfnamefont#1{#1}\fi
\expandafter\ifx\csname citenamefont\endcsname\relax
  \def\citenamefont#1{#1}\fi
\expandafter\ifx\csname url\endcsname\relax
  \def\url#1{\texttt{#1}}\fi
\expandafter\ifx\csname urlprefix\endcsname\relax\def\urlprefix{URL }\fi
\providecommand{\bibinfo}[2]{#2}
\providecommand{\eprint}[2][]{\url{#2}}

\bibitem[{\citenamefont{Caves}(1981)}]{Caves81}
\bibinfo{author}{\bibfnamefont{C.~M.} \bibnamefont{Caves}},
  \bibinfo{journal}{Phys. Rev. D} \textbf{\bibinfo{volume}{23}},
  \bibinfo{pages}{1693} (\bibinfo{year}{1981}).

\bibitem[{\citenamefont{Kok and Lovett}(2010)}]{kok10}
\bibinfo{author}{\bibfnamefont{P.} \bibnamefont{Kok}}  \bibnamefont{and}
  \bibinfo{author}{\bibfnamefont{B.~W.}~\bibnamefont{Lovett}},
  \emph{\bibinfo{title}{Introduction to optical quantum information processing}}
  (\bibinfo{publisher}{Cambridge University Press}, \bibinfo{year}{2010}).

\bibitem[{\citenamefont{{C. W. Helstrom}}(1967)}]{helstrom67}
\bibinfo{author}{\bibnamefont{{C. W. Helstrom}}}, \bibinfo{journal}{{Phys.
  Letters}} \textbf{\bibinfo{volume}{25A}}, \bibinfo{pages}{101}
  (\bibinfo{year}{1967}).

\bibitem[{\citenamefont{Helstrom}(1976)}]{helstrom76}
\bibinfo{author}{\bibfnamefont{C.~W.} \bibnamefont{Helstrom}},
  \emph{\bibinfo{title}{Quantum detection and estimation theory}}
  (\bibinfo{publisher}{Academic Press}, \bibinfo{year}{1976}).

\bibitem[{\citenamefont{Braunstein and Caves}(1994)}]{braunstein94}
\bibinfo{author}{\bibfnamefont{S.~L.} \bibnamefont{Braunstein}}
  \bibnamefont{and} \bibinfo{author}{\bibfnamefont{C.~M.} \bibnamefont{Caves}},
  \bibinfo{journal}{Phys. Rev. Lett.} \textbf{\bibinfo{volume}{72}},
  \bibinfo{pages}{3439} (\bibinfo{year}{1994}).

\bibitem[{\citenamefont{Gardiner and Zoller}(2004)}]{gardiner04}
\bibinfo{author}{\bibfnamefont{C.~W.} \bibnamefont{Gardiner}} \bibnamefont{and}
  \bibinfo{author}{\bibfnamefont{P.}~\bibnamefont{Zoller}},
  \emph{\bibinfo{title}{Quantum noise}}, \bibinfo{number}{p. 322}
  (\bibinfo{publisher}{Springer-Verlag}, \bibinfo{year}{2004}),
  \bibinfo{edition}{3rd} ed.

\bibitem[{\citenamefont{Holland and Burnett}(1993)}]{holland93}
\bibinfo{author}{\bibfnamefont{M.~J.} \bibnamefont{Holland}} \bibnamefont{and}
  \bibinfo{author}{\bibfnamefont{K.}~\bibnamefont{Burnett}},
  \bibinfo{journal}{{Phys. Rev. Lett.}} \textbf{\bibinfo{volume}{71}},
  \bibinfo{pages}{1355} (\bibinfo{year}{1993}).

\bibitem[{\citenamefont{Beltr{\'a}n and Luis}(2005)}]{luis05}
\bibinfo{author}{\bibfnamefont{J.}~\bibnamefont{Beltr{\'a}n}} \bibnamefont{and}
  \bibinfo{author}{\bibfnamefont{A.}~\bibnamefont{Luis}},
  \bibinfo{journal}{{Phys. Rev. A}} \textbf{\bibinfo{volume}{72}},
  \bibinfo{pages}{045801} (\bibinfo{year}{2005}).

\bibitem[{\citenamefont{{S. Boixo, S. T. Flammia, C. M. Caves, and J. M.
  Geremia}}(2007)}]{boixo}
\bibinfo{author}{\bibnamefont{{S. Boixo, S. T. Flammia, C. M. Caves, and J. M.
  Geremia}}}, \bibinfo{journal}{{Phys. Rev. Lett.}}
  \textbf{\bibinfo{volume}{98}}, \bibinfo{pages}{090401}
  (\bibinfo{year}{2007}).

\bibitem[{\citenamefont{Roy and Braunstein}(2008)}]{Roy08}
\bibinfo{author}{\bibfnamefont{S.}~\bibnamefont{Roy}} \bibnamefont{and}
  \bibinfo{author}{\bibfnamefont{S.~L.} \bibnamefont{Braunstein}},
  \bibinfo{journal}{Phys. Rev. Lett.} \textbf{\bibinfo{volume}{100}},
  \bibinfo{pages}{220501} (\bibinfo{year}{2008}).

\bibitem[{\citenamefont{Giovannetti et~al.}(2006)\citenamefont{Giovannetti,
  Lloyd, and Maccone}}]{Giovannetti06}
\bibinfo{author}{\bibfnamefont{V.}~\bibnamefont{Giovannetti}},
  \bibinfo{author}{\bibfnamefont{S.}~\bibnamefont{Lloyd}}, \bibnamefont{and}
  \bibinfo{author}{\bibfnamefont{L.}~\bibnamefont{Maccone}},
  \bibinfo{journal}{Phys. Rev. Lett.} \textbf{\bibinfo{volume}{96}},
  \bibinfo{pages}{10401} (\bibinfo{year}{2006}).

\bibitem[{\citenamefont{Breit and Wigner}(1936)}]{breit}
\bibinfo{author}{\bibfnamefont{G.}~\bibnamefont{Breit}} \bibnamefont{and}
  \bibinfo{author}{\bibfnamefont{E.}~\bibnamefont{Wigner}},
  \bibinfo{journal}{Phys. Rev.} \textbf{\bibinfo{volume}{49}},
  \bibinfo{pages}{519} (\bibinfo{year}{1936}).

\bibitem[{\citenamefont{Uffink}(1993)}]{uffink93}
\bibinfo{author}{\bibfnamefont{J.}~\bibnamefont{Uffink}}, \bibinfo{journal}{Am.
  J. Phys.} \textbf{\bibinfo{volume}{61}}, \bibinfo{pages}{935}
  (\bibinfo{year}{1993}).

\bibitem[{\citenamefont{Bhattacharyya}(1943)}]{bhatta43}
\bibinfo{author}{\bibfnamefont{A.}~\bibnamefont{Bhattacharyya}},
  \bibinfo{journal}{Bulletin of the Calcutta Mathematical Society}
  \textbf{\bibinfo{volume}{35}}, \bibinfo{pages}{99} (\bibinfo{year}{1943}).

\bibitem[{\citenamefont{Wootters}(1981)}]{wootters81}
\bibinfo{author}{\bibfnamefont{W.~K.} \bibnamefont{Wootters}},
  \bibinfo{journal}{Phys. Rev. D} \textbf{\bibinfo{volume}{23}},
  \bibinfo{pages}{357} (\bibinfo{year}{1981}).

\bibitem[{\citenamefont{Jones and Kok}(2010)}]{jones10}
\bibinfo{author}{\bibfnamefont{P.~J.} \bibnamefont{Jones}} \bibnamefont{and}
  \bibinfo{author}{\bibfnamefont{P.}~\bibnamefont{Kok}},
  \bibinfo{journal}{Phys. Rev. A} \textbf{\bibinfo{volume}{82}},
  \bibinfo{pages}{022107} (\bibinfo{year}{2010}).

\bibitem[{\citenamefont{Braunstein et~al.}(1996)\citenamefont{Braunstein,
  Caves, and Milburn}}]{braunstein96b}
\bibinfo{author}{\bibfnamefont{S.~L.} \bibnamefont{Braunstein}},
  \bibinfo{author}{\bibfnamefont{C.~M.} \bibnamefont{Caves}}, \bibnamefont{and}
  \bibinfo{author}{\bibfnamefont{G.~J.} \bibnamefont{Milburn}},
  \bibinfo{journal}{Ann. Phys.} \textbf{\bibinfo{volume}{247}},
  \bibinfo{pages}{135} (\bibinfo{year}{1996}).

\bibitem[{\citenamefont{Heitler}(1954)}]{heitler54}
\bibinfo{author}{\bibfnamefont{W.}~\bibnamefont{Heitler}},
  \emph{\bibinfo{title}{The Quantum Theory of Radiation}}, \bibinfo{number}{p.
  65} (\bibinfo{publisher}{Clarendon Press, Oxford}, \bibinfo{year}{1954}),
  \bibinfo{edition}{3rd} ed.

\bibitem[{\citenamefont{Margolus and Levitin}(1998)}]{margolus98}
\bibinfo{author}{\bibfnamefont{N.}~\bibnamefont{Margolus}} \bibnamefont{and}
  \bibinfo{author}{\bibfnamefont{L.~B.} \bibnamefont{Levitin}},
  \bibinfo{journal}{Physica D} \textbf{\bibinfo{volume}{120}},
  \bibinfo{pages}{188} (\bibinfo{year}{1998}).

\bibitem[{\citenamefont{Giovannetti et~al}(2003)}]{giovannetti03}
\bibinfo{author}{\bibfnamefont{V.}~\bibnamefont{Giovannetti}},
  \bibinfo{author}{\bibfnamefont{S.}~\bibnamefont{Lloyd}}, \bibnamefont{and}
  \bibinfo{author}{\bibfnamefont{L.}~\bibnamefont{Maccone}},
  \bibinfo{journal}{Phys. Rev. A} \textbf{\bibinfo{volume}{67}},
  \bibinfo{pages}{052109} (\bibinfo{year}{2003}).

\end{thebibliography}
\end{document}